\documentclass{article}
\usepackage{ijcai19}

\usepackage{times}
\usepackage{soul}
\usepackage{url}
\usepackage[utf8]{inputenc}
\usepackage[small]{caption}
\usepackage{graphicx}
\usepackage{amsmath}
\usepackage{booktabs}
\usepackage{multirow}
\usepackage{algorithm}
\usepackage{algorithmic}
\usepackage{pifont} 

\newcommand{\brm}[1]{\boldsymbol{\mathrm{#1}}} 
\newcommand{\ang}[1]{\langle#1\rangle} 
\def\b{\brm{b}}
\def\c{\brm{c}}
\def\H{\mathcal{H}}
\def\h{\brm{h}}
\def\m{\brm{m}}
\def\o{\brm{o}}
\def\t{\brm{t}}

\def\V{\brm{V}}
\def\W{\brm{W}}

\title{Patent Citation Dynamics Modeling via Multi-Attention Recurrent
       Networks}

\author{%
Taoran Ji$^{1,2}$\and
Zhiqian Chen$^{1,2}$\and
Nathan Self$^{1,2}$\and
Kaiqun Fu$^{1,2}$\and
Chang-Tien Lu$^{1,2}$\\\And%
Naren Ramakrishnan$^{1,2}$
\affiliations%
$^1$Discovery Analytics Center, Virginia Tech, Arlington, VA 22203, USA\\
$^2$Department of Computer Science, Virginia Tech, Arlington, VA 22203, USA\\
\emails%
\{jtr, czq, nwself, fukaiqun, ctlu\}@vt.edu,
naren@cs.vt.edu
}

\begin{document}

\maketitle

\begin{abstract} 
    Modeling and forecasting forward citations to a patent is a central task
    for the discovery of emerging technologies and for measuring the pulse of
    inventive progress.  Conventional methods for forecasting these forward
    citations cast the problem as analysis of temporal point processes which
    rely on the conditional intensity of previously received citations.  Recent
    approaches model the conditional intensity as a chain of recurrent neural
    networks to capture memory dependency in hopes of reducing the restrictions
    of the parametric form of the intensity function.  For the problem of
    patent citations, we observe that forecasting a patent's chain of citations
    benefits from not only the patent's history itself but also from the
    historical citations of assignees and inventors associated with that
    patent.  In this paper, we propose a sequence-to-sequence model which
    employs an attention-of-attention mechanism to capture the dependencies of
    these multiple time sequences.  Furthermore, the proposed model is able to
    forecast both the timestamp and the category of a patent's next citation.
    Extensive experiments on a large patent citation dataset collected from
    USPTO demonstrate that the proposed model outperforms state-of-the-art
    models at forward citation forecasting.
\end{abstract} 
\section{Introduction}%
\label{sec:introduction}

Patents are a direct outcome of research and development progress from industry
and as such can serve as a signal for the direction of technological advances.
Of particular interest are a patent's so-called forward citations: citations to
a patent from future patents.  Because new patents must cite related works and
those citations are vetted by the United States Patent and Trademark Office
(USPTO), forward citations serve as a marker of the importance of a patent.  As
a research topic, they have been used because: (1) they mirror global,
cumulative inventive progress~\cite{von2005inventive}; (2) they link
contributors and patents associated with the technology development cycle; (3)
they can discover emerging technologies at an early stage~\cite{lee2018early};
(4) they assist the measurement of patent quality~\cite{bessen2008value}; (5)
they contain information required for technology impact
analyses~\cite{jang2017hawkes}.  In today's competitive business environment,
citation forecasting is a field of growing importance.  By assessing the
long-term potential of new technologies, such forecasts can be instrumental for
decision making and risk assessment. Despite its importance as a proxy and
metric for many fields of technological impact study, forecasting an individual
patent's citations over time is a difficult task.  

The crux of current practice in this area is to treat the forward citation
chain of a patent as a temporal point process, governed by a conditional
intensity function whose parameters can be learned from observations of already
received citations.  Conventional methods for modeling temporal point processes
focus on modulating a specific parametric form of the intensity function under
strong assumptions about the generative process.  For example, a body of
studies on predicting ``resharing'' behaviors, such as forward citations of
papers and patents~\cite{xiao2016modeling,liu2017predictive} and
retweets~\cite{mishra2016feature}, built their models based on the Hawkes
process~\cite{hawkes1971spectra}.  In the Hawkes process, the intensity
function grows by a certain value whenever a new point arrives and decreases
back towards background intensity.  These approaches are limited by the
assumption that functional forms of conditional intensity will capture the
complicated dependencies among historical arrivals in real datasets.  Another
limit is that, with the Hawkes process, the probability density function, in
the form of maximum likelihood estimation, is computed by integrating the
conditional intensity function with respect to time.  Thus, due to the
complication of this computation, the choice of available decay kernels is
limited.  Throughout the literature, commonly used kernel decay functions are
variants of power-low
functions~\cite{helmstetter2002subcritical,mishra2016feature}, exponential
functions~\cite{xiao2016modeling,filimonov2015apparent,liu2017predictive} and
Reyleigh functions~\cite{wallinga2004different}.

More recent approaches~\cite{du2016recurrent,wang2017cascade,xiao2017modeling}
attempt to employ recurrent neural networks (RNN) to encode inherent dependency
structures over historical points. This removes the assumptions that the
generative process can be represented via a base intensity and a decay
function. RNN-based temporal point processes have been shown not only to
capture the intensity structure of synthetic data (simulated by typical point
process models such as Hawkes and self-correcting)~\cite{isham1979self}, but
also to outperform conventional methods on real world datasets. There are,
however, three drawbacks to exisitng RNN-based methods. First, in order to
simplify integration, intensity functions are usually represented as an
exponential function of the current hidden state of RNN units. In the training
stage, this can easily cause numerical instability when calculating the density
function for maximum likelihood estimation. And in the inference stage, this is
also a computation bottleneck.  Second, the goal of these methods is to predict
the arrival of the next point but for patent citations it is more important to
generate the entire sequence of subsequent points. Third, none of these methods
considers the interaction of multiple historical sequences.

In this paper, we propose a sequence-to-sequence model for patent citation
forecasting.  This model uses multiple historical sequences: one each for the
patent itself, its assignee, and its inventor.  These three are encoded
separately by RNNs and a decoder generates the citation forecast sequence based
on an attention-of-attention mechanism which modulates the intervening
dependency of the sequences embedded by the encoders.  Furthermore, instead of
explicitly designing the intensity function, we use nonlinear point-wise
feed-forward networks to generate the prediction directly.  Specifically, the
contributions and highlights of this paper are:
\begin{itemize}
    \item Formulating a sequence-to-sequence framework to jointly model patent
        citation arrival time and patent category information by learning
        a representation of observed historical sequences for the patent
        itself, its associated assignees, and related inventors.
    \item Designing an attention-of-attention mechanism which empowers the
        decoder to tackle the intervening dependency of these three historical
        sequences, which together contribute to the forward citation
        forecasting.
    \item Conducting comprehensive experiments on a large scale dataset
        collected from the USPTO to demonstrate that our model has consistently
        better performance for forecasting both the categories and the
        timestamps of new citations.
\end{itemize}

\section{Problem Formulation}%
\label{sec:problem_formulation}

For each target patent $p$, we compile three associated time-dependent
sequences. First, for the target patent itself, we collect a sequence of the
timestamp and category for existing forward citations $S_{p}
= \{(t_{p}^{\ang{1}}, m_{p}^{\ang{1}}), \ldots, (t_{p}^{\ang{k}},
m_{p}^{\ang{k}}), \ldots\}$ where $t_{p}^{\ang{k}}$ and $m_{p}^{\ang{k}}$ refer
to the time and the category of the forward citation, respectively.  Second, we
create a citation chain for the assignee who owns the target patent $S_{a}
= \{t_{a}^{\ang{1}}, \ldots, t_{a}^{\ang{k}}, \ldots\}$. This chain includes
patents that cite any patent owned by the assignee. Likewise, we compile the
sequence of citations of all patents invented by the inventor of the target
patent $S_{a} = \{t_{a}^{\ang{1}}, \ldots, t_{a}^{\ang{k}}, \ldots\}$. The key
idea of using these three sequences is that, in practice, we expect forecasts
to be made with a short observation window which means that the chain of
forward citations for the target patent may not have sufficient records for
training.

Given the input data as described above, our problem is as follows: for
a target patent $p$, taking the first $n$ points in $S_{p}$ as observations, as
well the historical citation records for its associated assignees
$\{t_{a}^{\ang{k}} | t_{a}^{\ang{k}} < t_{p}^{\ang{n}}\}$ and inventors
$\{t_{v}^{\ang{k}} | t_{v}^{\ang{k}} < t_{p}^{\ang{n}}\}$, can we predict at
what time this target patent will be cited by a new patent and to which
category that new patent will belong? The question breaks down into two
subproblems: 
\begin{enumerate}
    \item Predict the next citation
        $(t_{p}^{\ang{n + 1}}, m_{p}^{\ang{n + 1}})$ given the observed
        citation sequences for the patent, its assignee, and its inventor
        before time $t_{p}^{\ang{n + 1}}$,
    \item Predict the next $l$ citations $\{(t_{p}^{\ang{n + l}}, m_{p}^{\ang{n
        + l}})\}$ given the historical citation sequences of the patent, its
        assignee, and its inventor before time $t_{p}^{\ang{n + 1}}$.
\end{enumerate}

The first subproblem implies a many-to-many recurrent neural net where the
length of the input and output sequences matches the number of recurrent steps.
This first subproblem is easier to solve because at each prediction $k$ all
previous historical records before timestamp $t^{\ang{k}}$ are available. The
second subproblem is more meaningful for patent citation forecasting because it
requires the model to make the next $l$ predictions after timestamp
$t^{\ang{k}}$ based only on the observations from $t^{\ang{1}}$ to
$t^{\ang{k}}$. Our proposed model is focused on the second subproblem, and
consequently can also handle the first subproblem when $l = 1$. 

\section{Models}%
\label{sec:models}

\begin{figure*}[tpb]
    \centering
    \includegraphics[width=0.85\linewidth]{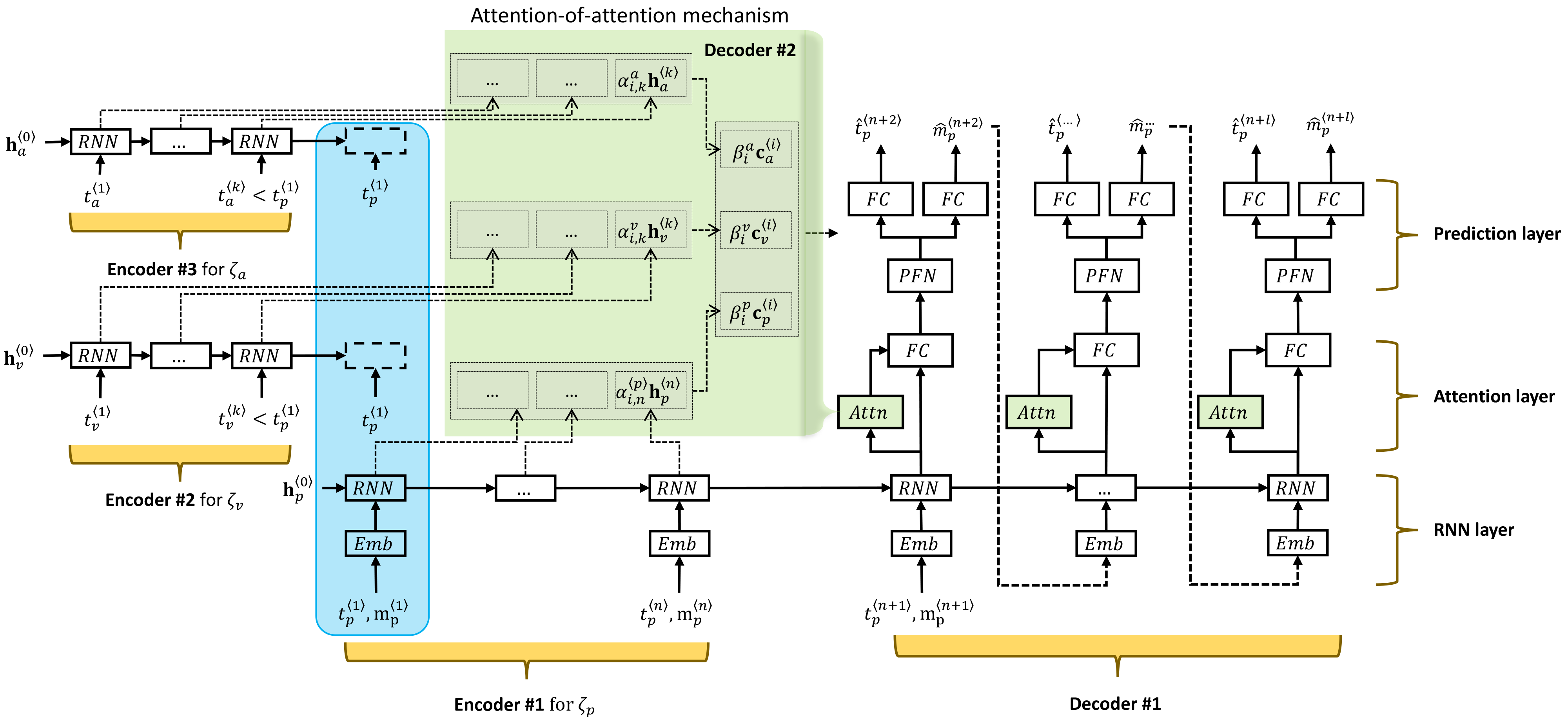}
    \caption{The architecture of PC-RNN.\ In addition to the patent
    citation chain which can only be observed during a short time window, the
    source-side RNN models take into account two other temporal processes which
    are assumed to play a critical role in modeling future citations of a
    patent: (1) the historical citations of the target patent's assignee
    managed by ``Encoder \#3'', and (2) the
    historical citations of the target patent's inventor encoded by
    ``Encoder \#2''. The target-side model consists of three sublayers: (1) the
    RNN layer which decodes the encoded sequence from source-side
    ``Encoder \#1'', (2) the attention layer which takes care about the
    intervening dependency structure across three sequences on source-side, and
    (3) the prediction layer which makes prediction for both the time and the
    category of the next forward patent citation.  
    }\label{fig:overview}
\end{figure*}

In this section, we present our proposed model, PC-RNN, which tackles the
problem of forecasting the next $l$ citations of a target patent. First, we
show an overview of the design of the proposed framework. Then we detail the
RNN architecture used on the source-side and the target-side of our model.
Next we introduce the attention-of-attention mechanism which plays an important
role in fusing multiple encoded sequences from the source-side for generating
the predicted citation sequence. Finally, we describe the training procedure
which unifies the distinct components of the proposed framework. 

\subsection{Model Overview}%
\label{sub:model_overview}

The chain of forward citations for a patent can be modeled as a marked temporal
point process where the joint distribution of all citations received for
a target paper $p$ can be described as the joint density
\begin{equation}
    \begin{aligned}
        f(S_{p}) & = \prod_{k}f\left((t_{p}^{\ang{k}}, m_{p}^{\ang{k}}) |
            \H^{\ang{k}}\right)\\
    \end{aligned}
\end{equation}
where $\H^{\ang{k}}$ denotes all the historical information associated with the
target patent at timestamp $k$, $t_{p}^{\ang{k}}$ is the timestamp when the
$k$-th citation arrives, and $m_{p}^{\ang{k}}$ is the category of the arriving
citation. Because our primary use case is to make predictions in the early
stages of a patent's life cycle, we compensate for having a short observation
window by incorporating citation sequences for the assignees and inventors
associated with the target patent. In particular, taking as inputs (1)
a time-dependent citation sequence of $n$ observations of the target patent
$\zeta_{p} = \{(t_{p}^{\ang{1}}, m_{p}^{\ang{1}}), \ldots, (t_{p}^{\ang{n}},
m_{p}^{\ang{n}})\}$, (2) records of citations received by associated assignees
$\zeta_{a} = \{t_{a}^{\ang{k}} | t_{a}^{\ang{k}} < t_{p}^{\ang{1}}\}$, and (3)
records of citations received by associated inventors $\zeta_{v}
= \{t_{v}^{\ang{k}} | t_{v}^{\ang{k}} < t_{p}^{\ang{1}}\}$, the task of the
proposed model is to learn an optimal distribution on observations:
\begin{equation}\label{equ:model}
    \Pr\left((t_{p}^{\ang{n + 1}}, m_{p}^{\ang{n + 1}}), \ldots,
        (t_{p}^{\ang{n + l}}, m_{p}^{\ang{n + l}})
        | \zeta_{p}, \zeta_{a}, \zeta_{v}\right),
\end{equation}
where $\zeta_{p}$, $\zeta_{a}$, and $\zeta_{v}$ together compose the observed
historical information.

Generally speaking, Eq.~\ref{equ:model} transforms the three given sequences
into another sequence. Thus, we are guided by the sequence-to-sequence
paradigm~\cite{sutskever2014sequence} to design the proposed framework. In the
overview of proposed model depicted in Fig.~\ref{fig:overview}, the module that
handles input sequences is called source-side and the module that makes
predictions is called target-side. On the source-side, three RNNs act as
encoders to separately encode the observations $\zeta_{p}, \zeta_{a},
\zeta_{v}$ which align at timestamp $t_{p}^{\ang{1}}$ when the first citation
of target patent arrives. In Fig.~\ref{fig:overview}, three input sequences
$\zeta_{p}, \zeta_{a}, \zeta_{v}$ are managed by ``Encoder \#1'',
``Encoder \#2'', and ``Encoder \#3,'' respectively. 
On the target-side, the decoder recurrently forecasts the time at which a new
patent will cite the target patent and in which category that new patent will
be. The target-side consists of three sublayers: a RNN layer, an attention
layer, and a prediction layer. In particular, the RNN layer decodes the patent
citation sequence from the source-side. The hidden state of the RNN layer
enters the attention layer where an attentional hidden state is computed by
fusing all the information embedded by the encoders. Finally, the prediction
layer makes category and timestamp predictions.  

\subsection{Ingredients of PC-RNN}%
\label{sub:ingredients_of_pc_rnn}

The three encoder RNNs separately handle the observed sequences $\zeta_{p},
\zeta_{a}, \zeta_{v}$ by compiling the inputs into an intermediate
fixed-dimensional hidden representation. As an example, we  describe the patent
sequence encoder to illustrate the computation process. For the 
patent sequence, at the $k$-th observation of $\zeta_{p}$, the category
information $m_{p}^{\ang{k}}$ of the citing patent is first projected into
a dense representation through an embedding layer and then fed into the RNN
along with temporal features: 
\begin{equation}\label{equ:encoder}
    \h_{p}^{\ang{k}} = \sigma_{p}\left(
        \W^{t}\t_{p}^{\ang{k}} + \W^{m}\m_{p}^{\ang{k}}
        + \W^{h}\h_{p}^{\ang{k - 1}} + \b^{h}\right),
\end{equation}
where $\t_{p}^{\ang{k}}$ represents the temporal features (e.g.,
$t_{p}^{\ang{k}} - t_{p}^{\ang{k - 1}}$), $\m_{p}^{\ang{k}}$ is a dense
representation of the category information, $\h_{p}^{\ang{k - 1}}$ is a
$d_{p}$-dimensional hidden state from the last step which embedded historical
information, and $\sigma_{p}$ is the activation function. For clarity, we use
the vanilla RNN in Eq.~\ref{equ:encoder} to illustrate the computation process.
In practice, we use multi-layered bidirectional LSTMs~\cite{hochreiter1997long}
which can capture long range temporal dependencies in sequences. In this case,
at each step $k$ the output of the encoder is the summation of the hidden state
of LSTMs from two directions: $\o_{p}^{\ang{k}}
= \overrightarrow{\h}_{p}^{\ang{k}} + \overleftarrow{\h}_{p}^{\ang{k}}$.  The
assignee sequence encoder for $\zeta_{a}$ and the inventor sequence encoder for
$\zeta_{v}$ share similar computational processes with the target patent
sequence encoder except that there is no category information to embed.  The
workflow of patent, inventor, and assignee encoders is depicted in
Fig.~\ref{fig:overview} as ``Encoder \#1'',
``Encoder \#2'', and ``Encoder \#3,'' respectively. 
On the target-side, the decoder RNN employs multi-layered unidirection LSTMs,
which adds the restriction that the input point at each step is always the
output from the last step.

\subsubsection{Attention Layer}%
\label{ssub:attention_layer}

Theoretically, the decoder layer generates a citation sequence by recurrently
maximizing the conditional probability
\begin{equation}
    \Pr\left(t_{p}^{\ang{i}}, m_{p}^{\ang{i}} | \h_{d}^{\ang{i}}, \zeta_{p},
        \zeta_{a}, \zeta_{v}\right), i > n,
\end{equation}
where $\h_{d}^{\ang{i}}$ is the hidden state of the decoder RNN at time step
$i$. However, an RNN layer alone cannot appropriately handle the 
conditional dependency over the three citation sequences
$\zeta_{p}, \zeta_{a}, \zeta_{v}$. Thus, we propose an attention layer on top
of the RNN layer. In particular, our model includes a two-level attention
mechanism~\cite{bahdanau2014neural,luong2015effective}. First, for each time
step $i$ in the decoder, a \textit{context} vector is computed for each encoder
RNN using all the encoder's outputs and the current hidden state of the decoder
$\h_{d}^{\ang{i}}$. Using outputs from patent encoder as an example, the
computation process is defined as:
\begin{equation}\label{equ:attn1}
    \begin{aligned}
        \c_{p}^{\ang{i}} &=
            \sum_{j \in \zeta_{p}}\alpha_{i, j}^{p}\o_{p}^{\ang{j}},\\
        \alpha_{i, j}^{p} & =
            \frac{\exp(e_{i, j})}{\sum_{k \in \zeta_{p}}\exp(e_{i, k})},\\
        e_{i, j} & = g(\h_{d}^{\ang{i}}, \o_{p}^{\ang{j}}) = \brm{V}^{p}\tanh(
        \brm{W}^{p}[\h_{d}^{\ang{i}}; \o_{p}^{\ang{i}}]),\\
    \end{aligned}
\end{equation}
where $\o_{p}^{\ang{j}}$ are encoder's outputs, $\alpha_{i, j}^{p}$ are the
attention weights, $g(\cdot)$ is an alignment function calculating the
alignment weights between position $j$ of the encoder and the $i$-th position
of the decoder, and $\brm{V}^{p}, \brm{W}^{p}$ are parameters to learn. Of the
four alignment functions provided in~\cite{luong2015effective}, we use the
``concat'' function here because, in our case, source-side encoders output
multiple sequences of different dimensions than the decoder's hidden state. In
our experiments, the dimension of the hidden state of the patent sequence
encoder and decoder is set to 32 and is 16 for assignee and inventor sequence
encoders. In this step, the decoder can learn to attend to certain units of
each encoder which enables the model to learn complicated dependency
structures. Next, the attention weights for each context vector are calculated,
based on which weighted context vectors are concatenated as the input to the
next layer:
\begin{equation}\label{equ:attn2}
    \begin{aligned}
        \bar{\c}^{\ang{i}} & = [\beta_{i}^{p}\c_{p}^{\ang{i}};
                               \beta_{i}^{a}\c_{a}^{\ang{i}};
                               \beta_{i}^{v}\c_{v}^{\ang{i}}],\\
    \end{aligned}
\end{equation}
where the calculation of $\beta$ is similar to the calculation of
$\alpha$ in Eq.~\ref{equ:attn1}. We perform this additional attention
calculation because we want the model to dynamically learn and determine the
combination of context vectors from each encoder which should enhance the
flexibility of the modeled memory dependency structure. We call this two-step
operation an attention-of-attention mechanism.  It is depicted in the green
rectangle labeled ``Decoder \#2'' in Fig.~\ref{fig:overview}. Finally, the
concatenated context vector $\bar{\c}^{\ang{i}}$ is combined with hidden state
$\h_{d}^{\ang{i}}$ to generate the attentional hidden state
$\bar{\h}_{d}^{\ang{i}}$ which flows to the next prediction layer:
\begin{equation}
    \begin{aligned}
        \bar{\h}_{d}^{\ang{i}} & = \phi\left(
            \V^{c}[\bar{\c}^{\ang{i}};\h_{d}^{\ang{i}}]\right),\\
    \end{aligned}
\end{equation}
where $\phi$ is a non-linear activation function, and $\V^{c}$ is the parameter
to learn. In our experiments, we use ReLU activation function.

\subsubsection{Prediction Layer}%
\label{ssub:prediction_layer}

In the prediction layer, for each position in the attentional hidden state
$\bar{\h}_{d}^{\ang{i}}$, two linear transformations are applied with a
non-linear activation $\varrho$ in between to further enhance the model's
capability:
\begin{equation}
    \mathrm{PFN}(h) = w^{h2}\varrho\left(w^{h1}h + b^{h1}\right)
        + b^{h2},
\end{equation}
where $h$ is one position in $\bar{\h}_{d}^{\ang{i}}$, and $w^{h1}, w^{h2},
b^{1}, b^{2}$ are parameters to learn. The output of this layer is denoted by
$\tilde{\h}_{d}^{\ang{i}}$. This operation was introduced
in~\cite{vaswani2017attention} which is equivalent to two $1$-by-$1$
convolutions which is also called the ``network in network'' concept
in~\cite{lin2013network}. In our experiments, $\varrho$ is a ReLU function.
Finally, we use two fully connected layers to generate predictions:
\begin{equation}
    \begin{aligned}
        \hat{t}^{\ang{i + 1}} & = \max\left(
            \W^{t}\tilde{\h}_{d}^{\ang{i}} + b^{t}, 0\right),\\
        \hat{\m}^{\ang{i + 1}} & = \mathrm{softmax}\left(
            \W^{m}\tilde{\h}_{d}^{\ang{i}} + b^{m}\right).\\
    \end{aligned}
\end{equation}

The total loss is the sum of the time prediction loss and the cross-entropy
loss for the patent category prediction: 
\begin{equation}
    \begin{aligned}
        \mathcal{L} & = -\sum_{j=1}^{l}\left(
            \log\Pr((t_{p}^{\ang{n + j}}, m_{p}^{\ang{n + j}}))\right)\\
        & = -\sum_{j=1}^{l}\left(
            \log{\hat{\m}_{\kappa}^{\ang{n + j}}}
            + \vartheta(\hat{t}^{\ang{n + j}}, t^{\ang{n + j}})
            \right),
    \end{aligned}
\end{equation}
where $n$ refers to the number of points in the sequence used as observations,
$l$ is the number of predictions to make, $\kappa = m^{\ang{n + j}}$ indexes
the target category and consequently $\hat{\m}_{\kappa}^{\ang{n + j}}$ is the
probability of predicting the correct category in prediction $\m^{\ang{n
+ j}}$, and $\vartheta(\cdot)$ is defined as the absolute distance between
$\hat{t}^{\ang{n + j}}$ and $t^{\ang{n + j}}$. We adopted the
ADAM~\cite{kingma2014adam} optimizer for training.

\section{Experiments}%
\label{sec:experiments}

We compare our PC-RNN experimentally to state-of-the-art methods for modeling
marked temporal point processes on a large patent citation dataset collected
from USPTO.\ The results show that PC-RNN outperforms the others at modeling
citation dynamics.

\subsection{Dataset Description and Experiment Setup}%
\label{sub:dataset_description_and_experiment_setup}

\textbf{Dataset}: Our dataset originates from the publicly accessible
PatentsView\footnote{http://www.patentsview.org/download/} database which
contains more than $6$ million U.S.\ patents. According to patent law, new
inventions must cite prior arts and differentiate their innovations from them.
Patent citations have a high quality because citations are examined by
officials before patents are granted. For each patent, we construct a forward
citation chain using timestamps from the database. We remove patents with
citation chains shorter than 20 or longer than 200. Long citation chains are
relative rare in practice and the unbalanced sequence length may make training
difficult. For each citing patent we also extract its NBER~\cite{hall2001nber}
category information. As a result, each patent has one main category and one
subcategory. In total, there are 7 main categories and 37 subcategories in the
entire patent database. For each patent, we also retrieved the associated
assignees and inventors from the database and compiled the related citation
chains for each. For patents with multiple assignees or inventors, we selected
the longest available chain to provide the most information. Finally, having
assembled the dataset, we sampled 15,000 sequences of which 12,000 sequences
are used for the training set and the remaining 3,000 sequences are the test
set. Our dataset and code is publicly available for
download.\footnote{https://github.com/TaoranJ/PC-RNN}

\textbf{Metrics}: Following the similar procedure
in~\cite{du2016recurrent,wang2017cascade,xiao2017modeling}, two metrics are
used to compare the performance of different tasks. For event time prediction, 
we used \textit{Mean Absolute Error} (MAE). For both main category and
subcategory prediction, we used the \textit{Accuracy} metric. 

\textbf{Comparison Models:} State-of-the-art RNN based prediction methods can
be generally categorized into two classes. (1) Intensity-based RNNs represents
an observed sequence with the hidden state of a recurrent neural network. They
then formulate an intensity function conditioned on the hidden state to make
predictions as in a typical temporal point process. (2) End-to-end RNNs avoid
explicitly designated conditional intensity functions, and instead represent
the mapping from input sequences to predictions implicitly within the network
structure. Our method belongs to the second class. In our experiments, we
compare our model against three other major peer recurrent temporal point
process models of both classes, each of which predicts both the timestamp and
category of arrival points using observed sequences as input:
\begin{itemize}
    \item RMTPP~\cite{du2016recurrent}: Recurrent marked temporal point process
        (RMTPP) is an intensity-based RNN model for general point process
        analysis which is able to predict both the timestamp and the type of
        point in an event sequence. RMTPP is not a sequence-to-sequence model,
        however, by directing the $k$-th output to the $(k+1)$-th input, the
        model can be used as a sequence generator. RMTPP only models one
        sequence. So, in the experiment, we feed only the patent sequence
        to the model as input. 
    \item CYAN-RNN~\cite{wang2017cascade}: CYAN-RNN is an intensity-based RNN
        model which models a general resharing dependence in social media
        datasets. It can forecast the time and user of the next resharing
        behavior. Similarly to RMTPP, by forcing the $k$-th output to the
        $(k+1)$-th input, this model is used as a sequence generator for patent
        citation prediction. For this experiment, we use only patent
        citation sequences as input to generate predictions.
    \item IntensityRNN~\cite{xiao2017modeling}: In IntensityRNN, a
        time-dependent event arrival sequence and an evenly distributed
        background time series work together to generate predictions for the
        next arrival event. In experiments, the evenly distributed background
        time series is simulated with assignee citations which usually 
        have a stable citation pattern. IntensityRNN is an end-to-end RNN
        model.
\end{itemize}

\begin{table}[tpb]
    \centering
    \begin{tabular}{@{}lllll@{}}
    \toprule
        \multicolumn{5}{c}{$80\%$ As Observations} \\\midrule
        \multirow{2}{*}{Model} & \multicolumn{2}{c}{Main-category}
        & \multicolumn{2}{c}{Sub-category} \\\cmidrule(l){2-3}\cmidrule(l){4-5}
            & ACC & MAE & ACC & MAE \\ \midrule
        RMTPP           & 0.2316 & 0.0550 & 0.1213 & 0.0545 \\
        CYAN-RNN        & 0.2304 & 0.0566 & 0.1209 & 0.0546 \\
        IntensityRNN    & 0.6826 & 0.0218 & 0.5258 & 0.0220 \\
        PC-RNN          & \textbf{0.9498}  & \textbf{0.0216} & \textbf{0.7885}
                        & \textbf{0.0216}\\\midrule
        \multicolumn{5}{c}{$50\%$ As Observations} \\\midrule
        \multirow{2}{*}{Model} & \multicolumn{2}{c}{Main-category}
        & \multicolumn{2}{c}{Sub-category} \\\cmidrule(l){2-3}\cmidrule(l){4-5}
            & ACC & MAE & ACC & MAE \\ \midrule
        RMTPP           & 0.1743 & 0.1310 & 0.0761 & 0.1320 \\
        CYAN-RNN        & 0.2376 & 0.1055 & 0.1244 & 0.0987 \\
        IntensityRNN    & 0.6704 & 0.0178 & 0.5141 & 0.0177 \\
        PC-RNN          & \textbf{0.7885} & \textbf{0.0172}
                        & \textbf{0.6752} & \textbf{0.0172}\\ \midrule
        \multicolumn{5}{c}{$30\%$ As Observations} \\\midrule
        \multirow{2}{*}{Model} & \multicolumn{2}{c}{Main-category}
        & \multicolumn{2}{c}{Sub-category} \\\cmidrule(l){2-3}\cmidrule(l){4-5}
            & ACC & MAE & ACC & MAE \\ \midrule
        RMTPP           & 0.2404 & 0.1536 & 0.1292 & 0.2317\\
        CYAN-RNN        & 0.2434 & 0.1255 & 0.1269 & 0.1310\\
        IntensityRNN    & 0.6652 & \textbf{0.0163} & 0.4928 & 0.0167\\
        PC-RNN        & \textbf{0.7778} & 0.0165 & \textbf{0.6631}
                        & \textbf{0.0164}\\
        \midrule
        \multicolumn{5}{c}{$10\%$ As Observations} \\\midrule
        \multirow{2}{*}{Model} & \multicolumn{2}{c}{Main-category}
        & \multicolumn{2}{c}{Sub-category} \\\cmidrule(l){2-3}\cmidrule(l){4-5}
            & ACC & MAE & ACC & MAE \\ \midrule
        RMTPP           & 0.1769 & 0.3383 & 0.1299 & 0.4606\\
        CYAN-RNN        & 0.2443 & 0.1534 & 0.1296 & 0.1533\\
        IntensityRNN    & 0.6701 & 0.0182 & 0.4795 & \textbf{0.0167}\\
        PC-RNN          & \textbf{0.7362} & \textbf{0.0169} & \textbf{0.6331}
                        & 0.0169\\
        \bottomrule
    \end{tabular}
    \caption{Performance evaluation of our method and peer
        methods. Timestamp predictions are evaluated using MAE.\ Patent
         category predictions are evaluated using
        ACC.\ }\label{tab:performance}
\end{table}

\subsection{Results and Discussion}%
\label{sub:results_and_discussion}

The motivation for our model is to use the historical forward citations of
a patent's assignees and inventors to boost the prediction performance since
the observation window of the target patent is limited.  Therefore, our
experiment tests observation windows of different lengths.  In particular, in
four experiments, we use $80\%$, $50\%$, $30\%$ and $10\%$ of the patent
sequence as observations.  As the observation window becomes shorter, the
predictive task becomes more difficult but also more salient for real world
applications.  In each experiment, to fully examine each model's performance at
category prediction, each model runs the following two tasks separately:
\begin{itemize}
    \item prediction of the timestamp and the NBER main category of the newly
        arrived citation. 
    \item prediction of the timestamp and the NBER subcategory of the newly
        arrived citation. 
\end{itemize}

Experiments are conducted with the test dataset which is separate from the
training set.  Results are reported in Table~\ref{tab:performance}.  We
consider RMTPP and CYAN-RNN to be ``intensity-based RNNs'' because, instead of
using recurrent neural nets, they treat hidden states as the embedded
historical information from which they directly predict using an explicitly
conditional intensity function. In contrast, IntensityRNN belongs to the
end-to-end RNN model because it directly makes desired predictions given input.
To better understand the experimental results, we analyze the performance of
each class as compared to our model.  

\subsubsection{Intensity-Based Models}%

Our model consistently and significantly outperforms RMTPP and CYAN-RNN for
category classification in all experiments.  This suggests that these two
models struggle to modulate patent categories and subcategories.  Similar
results about weak classification performance using these two models on
different datasets are reported in~\cite{du2016recurrent,wang2017cascade}.  For
prediction of the next forward citation, our model consistently outperforms
these intensity-based models.  As the observation window shrinks, the
performance of our model on the category classification task decreases, though
still outperforms its competitors.  However, the performance of our model on
the time prediction task stays relatively stable.  We observe that the
timestamp information contained in the assignee's and inventor's citation
chains could help to predict upcoming citations to the target patent.  In
contrast, the performance of the intensity-based models on the time prediction
task decreases significantly as the available observations are reduced.  We
observe that the optimizations for intensity-based methods is sensitive to the
number of observations, especially considering that maximum likelihood
estimation is used. As a result, we conclude that including sequences
associated to the main target sequences in our model helps to improve its
sensitivity to the number of observations.

\subsubsection{End-to-End Models}%

Our model consistently outperforms IntensityRNN on the category classification
task.  In particular, for main category classification, our model is $6\%$ to
$26\%$ more accurate compared to IntensityRNN across all observation windows.
This improvement is even more significant for subcategory classification, where
our model has an improvement of $16\%$ to $26\%$. We argue that this boost is
attributed to our attention-of-attention mechanism. In our mechanism, the
first attention layer allows the decoder to look back at all the hidden states
for each encoder at each step and dynamically attend to only the important
states in each sequence. The second attention layer empowers the model to fuse
attended parts from different sequences which manages the dependencies across
input sequences. In contrast, IntensityRNN relies solely on the last hidden
state to carry the entire sequence's information and the two input sequences
are connected only through a fully connected layer which may cause information
loss. On the time prediction task, our model outperforms IntensityRNN on most
test cases, with the exception of two.  Considering IntensityRNN's poorer
performance at the category classification task in those two tests, we suggest
that IntensityRNN's loss functions focus more on time loss which impacts
classification performance.  In contrast, the loss function in our model is
more balanced and encourages the model to pay attention to the two tasks
equally.

In summary, we conclude that incorporating assignee and inventor information in
the source-side encoder provides extra information which improves our model's
performance at the prediction task.  Additionally, as a proxy for prediction,
using an end-to-end learning diagram to modulate a marked temporal point
process is better than using a specific parametric form of a conditional
intensity function, at least in the domain of patent citation classification.
Finally, the empirical results suggest that the attention-of-attention
mechanism enables the decoder to dynamically observation historical information
from different input sequences and consequently enhance model performance.  

\section{Conclusion}%
\label{sec:conclusion}

In this paper, we present a patent citation dynamics model with fused
attentional recurrent network. This model expands the traditional
sequence-to-sequence paradigm by considering multiple sequences on the
source-side: one each for the target patent, its assignee, and its inventor.
We employ an attention-of-attention mechanism which encourages the decoder to
attend to certain parts of each source-side sequence and then further fused the
attended parts, grouping the by sequences which leverage dynamically learned
weights. Our model is evaluated on on patents data collected from USPTO and
experimental results demonstrate that our model can consistently outperform
state-of-the-art temporal point process modeling methods at predicting time of
newly arrived citations and classification of their category and subcategory.


\section*{Acknowledgments}
This work is supported in part by the National Science Foundation via grants
DGE-1545362 and IIS-1633363. The US Government is authorized to reproduce and
distribute reprints of this work for Governmental purposes notwithstanding any
copyright annotation thereon. Disclaimer: The views and conclusions contained
herein are those of the authors and should not be interpreted as necessarily
representing the official policies or endorsements, either expressed or
implied, of NSF, or the U.S. Government.
\clearpage

\bibliographystyle{named}
\bibliography{ijcai19}

\end{document}